\documentclass[acmtoit,acmnow]{acmtrans2m}

\usepackage{url,epsfig}

\title{Trading Agents for Roaming Users}

\author{MAGNUS BOMAN, MARKUS BYLUND and FREDRIK ESPINOZA\\ Swedish Institute of Computer Science (SICS) AB
	\and MATS DANIELSON and DAVID LYB\"ACK \\ Financial Market Systems, OM AB}
            
\begin{abstract}
Some roaming users need services to manipulate autonomous processes. Trading agents running on agent trade servers are used as a case in point. We present a solution that provides the agent owners with means to upkeeping their desktop environment, and maintaining their agent trade server processes, via a briefcase service.
\end{abstract}

            
            
            
\begin{document}
            
            
\maketitle

\section{Trading Agents}
\label{sec:ta}
Unresolved questions related to law, ethics, security, and policy~\cite{egzi00} do not in general prevent stock markets from applying open policies to program trading~\cite{nasd99}. A \emph{trading agent} is a piece of encapsulated autonomous software that codes the preferences of its owner. In theoretical research, trading agents have been used in idealized games~\cite{RuMiPa94}, artificial markets (see, e.g., \url{http://ArtStkMkt.sourceforge.net/}), and competitions (such as the Trading Agent Competition, see \url{http://www.sics.se/tac/}). 

Human owners of trading agents have different levels of involvement in the coding itself. An expert agent owner writes her own code. A less experienced agent owner might employ a broker, which will write and carry responsibility for the code. An unexperienced agent owner can, on some markets, select an agent from a set of templates, such as a greedy agent, a value investor agent, a trend investor agent, etc. Once the agent has been written, it must be checked with respect to the rules of the marketplace. If the agent is to run on a server, i.e. if the marketplace supports an \emph{agent trade server}, the agent code must comply with the API and with traffic regulations, such as the use of proxies. The agent trade server will usually complement an electronic exchange system, in such a way that the core services are unaffected, and should~\cite{lybo01b}:
\begin{itemize}
\item open up for proper best-effort transport connectivity, fully in line with Internet design foundations and contemporary trends in the industry
\item prepare for the introduction of many new order details and tactical trading instructions
\item provide a new level of containment to better defend the complex system from fatal software error propagation when new services are introduced
\end{itemize}

Once the agents are up and running on the server, the owner will require straightforward means to modifying, maintaining, and even manipulating her agents. She can use various authenticated ways of reporting an update of her preferences, such as encrypted mail, https-server commands, telephone calls with PIN-code login procedures, etc. Such maintenance commands can be directed to the marketplace owner or to a broker, but regardless there will be rules and perhaps costs for the changes made. In the other direction, the owner will require reports on how her agent is doing. It is pivotal that this data is disseminated with as little delay as possible, as this data carries the information that might prompt changes to the agent, or changes to the preferences of the owner.

We seek to complicate the picture further by focussing on roaming agent owners. In section~\ref{sec:roam}, we describe the problem and background to trading on-the-fly. In section~\ref{sec:sol}, we describe the sView architecture that to some extent provides the solution. Since this is work in progress, we close with directions for future research.

\section{Roaming Trading Agent Owners}
\label{sec:roam}
Agent trading in theory relaxes the agent owner from monitoring market data, but in practice there will be long periods of intense trading, radical market changes, preference changes, and other factors that will contribute to the need for full control. Hence, the agent owners cannot afford to be stationary, or even what is normally referred to as mobile users. Instead they will be carrying with them the trading services, and will expect at all times at least one channel of swift interaction means. In practice, this could mean that the owner is carrying with her a portable computer with a wireless LAN-card, together with all the necessary subscriptions and payphone cards for use of hot spots. On that computer, a service in the form of a running computer program will be activated. 

A benefit of having trading agents is the possibility to express higher-level intentions, rather than explicit low-level orders. For stocks or commodities, this includes repositioning a portfolio with respect to sectors, shifting from stocks to bonds or vice versa, or acting on macro-economic information or news flashes. To find the best venue for a trade, several market places must be monitored in advance to find the optimal combination of order book and liquidity, and optimizing only one of the parameters will not suffice (cf.~\cite{OEO96}). This is no trivial task, since varying trading rules might have hard-to-foresee complications for the deal execution~\cite{MaOh97}. For large deals, you would want the company's most experienced trader to assist the agent, regardless of her physical location at the time of the deal.

In the future, a work group collectively in charge of some important asset class will need to dispatch responsibility for actions based on availability. The availability should be ranked based on bandwidth available now and a forecast for the coming period, as a complement to workforce capacity. Traditional work flow models seem not to account for this aspect of roaming. For instance, bi-directional roaming access over several media may be required and as long as access networks are heterogeneous and bandwidth constrained, an availability map must be maintained. The best candidates for support of roaming agent owners are currently found among personal service provision systems, where personal and group states and information is preserved and maintained.

\section{The {s}View System}
\label{sec:sol}

The sView System~\cite{ByEs00} is among other things an environment for personal services~\cite{By01}, some of which may be agents. Detailed descriptions can be found at the system home page~\url{http://sview.sics.se}. The system assumes a client/server model, with server access being channeled through a virtual service briefcase. The briefcase in turn supports access from many different types of devices and user interfaces. It is also private to an individual user, and it can store service components containing both service logic and data from service providers. This allows the service provider to split the services in two parts. One part provides commonly used functionality and user-specific data that executes and is stored within the virtual briefcase. The other part provides network-based functionality and data that is common to all users. Finally, the service briefcase is mobile and it can follow its user from host to host. This allows local and partially network-independent access to the service components in the briefcase.

The core sView specification constitutes an API that defines the basics of service components and personal service environment handling. It specifies service components, a runtime environment, a data structure for persistent and mobile images of service environments (service briefcases), and a server for handling service briefcases. The specification has been implemented as a set of Java packages. 
Service components can implement an arbitrarily large part of the functionality of the service, and range from mere proxies to web-based services, to standalone services. They can be declared persistent and/or mobile. A \emph{persistent} service component can save its state together with the service environment when the environment is saved locally on a host or migrates to another host in the network. If a service component is \emph{mobile} it will follow the service environment as it migrates to a different host in the network. 

The entity maintaining and providing runtime support to service components is the service context. Once the service components are loaded, the service context controls and sets the states of the service components. For example, newly created service components should be taken through an initialization state and when the service environment is about to migrate, all service components should be suspended. The service context is responsible for mediating service interfaces between service components. It is straightforward to implement a message passing service component on top of the core sView service/service communication scheme, and the sView reference implementation includes such a service. Via the service context, service components can control the behavior of the service environment (e.g., migration and shutdown) as well as the behavior of other service components (such as creation, suspension, and resumption). However, since these activities are sensitive matters, the user must grant privileges to each service component in order for them to perform these actions. A service component may for example be granted the privileges to create and load, but not suspend or kill, service components within the environment. Another service component may be allowed to initiate a migration of the whole service environment to another host, but not to control any aspect of individual service components.

A service briefcase server specifies an API for service briefcase handling. It includes basic functionality such as create new, get, put, and delete briefcase. It also includes functionality for synchronization of content in different instances of a service briefcase on different servers. Synchronization is used when a service briefcase is to be moved to a server on which it has already been stored. In this case it is possible that parts of the briefcase, such as service component specifications, need not be sent with the briefcase. Synchronization is performed in two steps: the first step is to find out the difference between the two service briefcases followed by the second step to update the parts that have been changed. The two-phase commit protocol is used to ensure data integrity during synchronization.

\begin{figure}
\epsfig{file=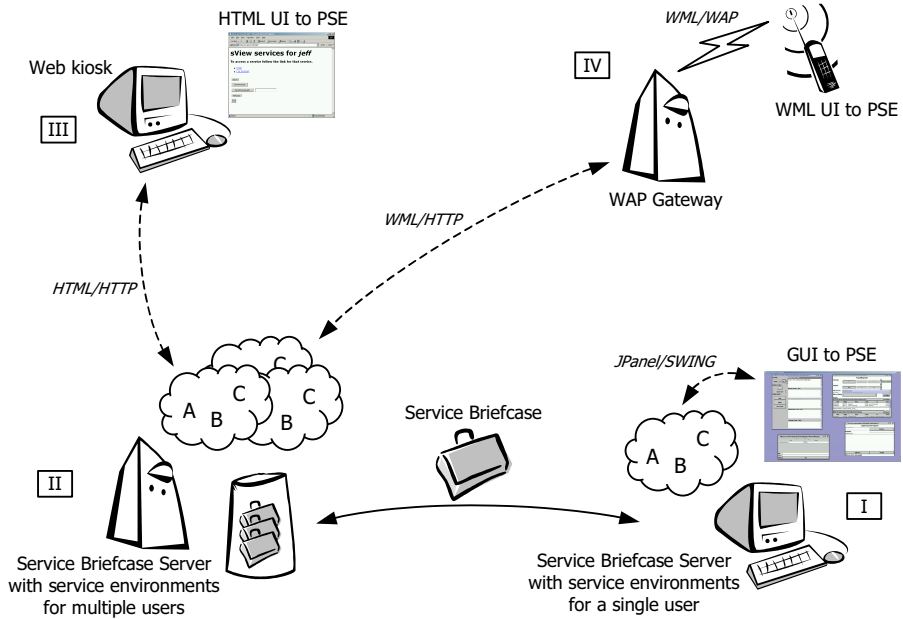}
\caption{Parts of the core sView specification and its APIs.}
\label{sviewfig}
\end{figure}

An illustration of the different parts of the core sView specification and their relations is given in Fig.~\ref{sviewfig}. On the computer marked I a briefcase server and a service environment are executing. In this case the user is sitting next to the same computer as the service environment (represented by the cloud together with service components A, B, and C) is executing on. This makes it possible to use a standard GUI for user-service interaction. The computer marked II hosts service briefcases and environments for several users, which use remote interfaces. One user is using a Web-kiosk with a Web browser for user-service interaction (III) and another user uses a WAP phone (IV). Stored service environments, in the form of service briefcases (illustrated between I and II), can migrate between any computers that run a briefcase server.

\section{Conclusions and Further Research}
The sView system is a state-of-the-art tool for letting users develop and maintain their electronic services. In particular, the system may be used as a tool for developing and maintaining services design to, in turn, monitor and manipulate agents executing on an agent trade server. An example of such a service is graphical illustrations of trades conducted by an agent, or by a team of agents. Another example is a tool for kill requests: should agent owner preferences change during the trading day, means to remove an agent entirely from the agent trade server must be available. Whether this privilege is granted the agent owner, or only member brokers for instance, is up to the marketplace owner to decide.

The sView system does not solve all problems resulting from roaming traders. For example, the issue of hot login must be solved. A roaming trader wishes independence from service providers, security mechanisms, and billing procedures. A login, presumably at a predefined priority level, should be sustained for as long as possible, and with graceful degradation if necessary. We are currently looking deeper into the case of sView being employed for roaming traders, and have begun to write actual services.

We have also started work on the actual agent trade server specification and code. Since our long-term goal is to generalize our experiences from this case to other domains, constructing a sandbox environment for trading agent development, which can later be reused, is also on our agenda.

\bibliographystyle{acmtrans}
\bibliography{mrt4}

\end{document}